# Curated loci prime editing (cliPE) for accessible multiplexed assays of variant effect (MAVEs)

**Carina G Biar,[1,2,3] Nicholas Bodkin[1,3], Gemma L Carvill[1,4], Jeffrey D Calhoun[1,4,5,6,*]**

[1]Ken and Ruth Davee Department of Neurology, Northwestern Feinberg School of Medicine, Chicago, Illinois, 60611 USA
[2]Genome Sciences, University of Washington, Seattle, Washington, 98195 USA
[3]Indicates equal contribution
[4]Indicates senior authorship
[5]Technical contact
[6]Lead contact
*Correspondence: jeffrey.calhoun@northwestern.edu

## Summary
Multiplexed assays of variant effect (MAVEs) perform simultaneous characterization of many variants. Prime editing has been recently adopted for introducing many variants in their native genomic contexts. However, robust protocols and standards are limited, preventing widespread uptake. Herein, we describe curated loci prime editing (cliPE) which is an accessible, low-cost experimental pipeline to perform MAVEs using prime editing of a target gene, as well as a companion Shiny app (pegRNA Designer) to rapidly and easily design user-specific MAVE libraries.

# Graphical abstract

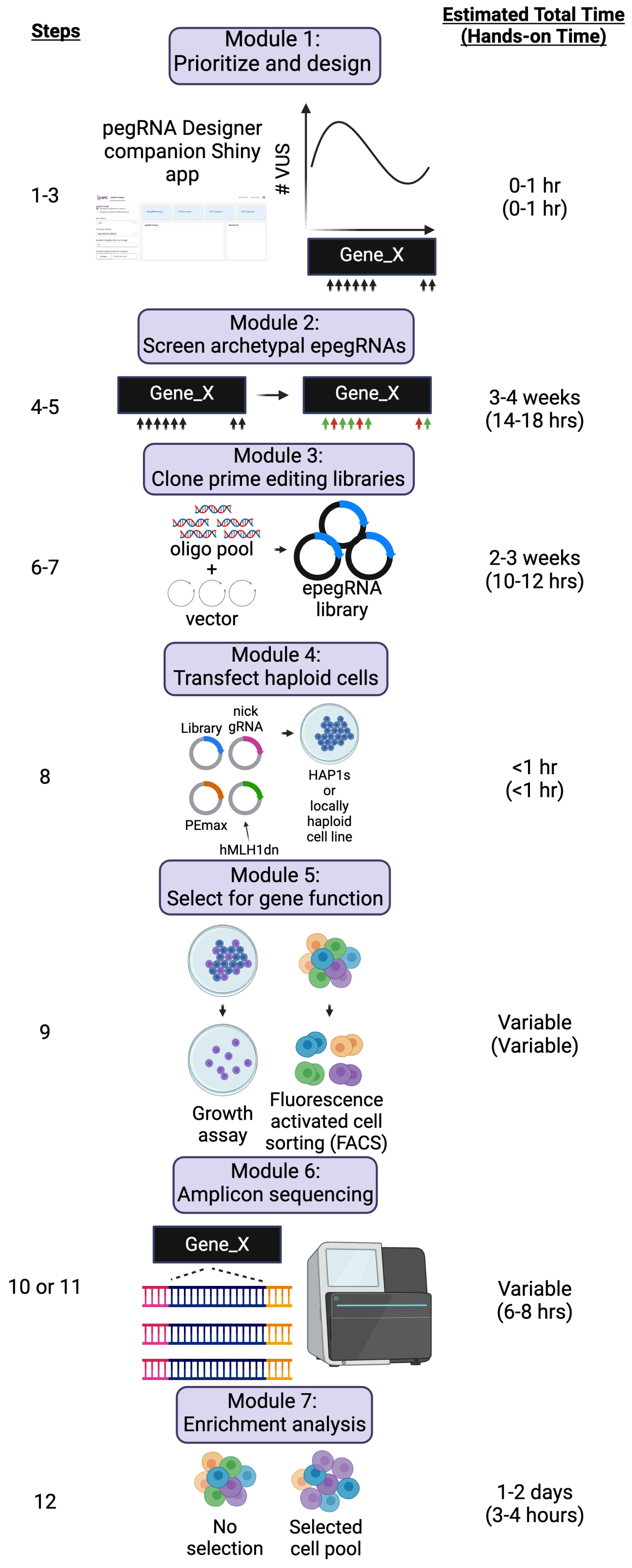

Clinical genetic testing has become one of the first-line diagnostic tests in modern medicine, spanning many specialties. As more and more variants causing genetic disorders are identified, our diagnostic yield continues to improve. However, there is a significant bottleneck which precludes precision genetic diagnostics: variants of uncertain significance (VUS). VUS occupy the grey area between pathogenic and benign variants and are often missense variants, or single amino acid substitutions, which may or may not impact protein function. The number of VUS reported per year is consistently increasing in the ClinVar database, the most comprehensive database of genetic testing results currently available to researchers, genetic counselors, clinicians, and the public. In fact, there are over 1 million VUS present in ClinVar as of 2024.[1] This steady increase in reported VUS has outpaced the scalability of classic, low-throughput functional assays of variant effect. High-throughput multiplexed assays of variant effect (MAVEs) are a relatively new class of technologies which leverage a library of genetic variants, a selection assay based on function of a particular gene, and next-generation sequencing (NGS) to collectively generate functional data on many variants simultaneously. MAVEs are an attractive option to address the VUS problem as they can scale to assess hundreds to thousands of variants, a much higher throughput than previous strategies.

Variant libraries in MAVEs can be made from expression plasmids or genome engineering of endogenous loci. Recently, Erwood S et al. developed saturation prime editing (SPE) which uses prime editing to generate a library of variants for the MAVE.[2] Prime editing is a modification of CRISPR/Cas9 genome editing which does not require a double-stranded DNA break.[3] Instead, prime editing is performed by co-expressing at least two components in a cell: (1) a nicking Cas9 fused to a reverse transcriptase and (2) a prime editing guide RNA (pegRNA), which contains both a spacer for targeting and a reverse transcription template to generate a variant of interest. While typical prime editing experiments use a single pegRNA to generate a single variant, Erwood et al. substituted a pegRNA library to generate a cell pool where each individual cell expresses one variant from the library.[2] A number of modifications to basic prime editing have improved its efficiency and therefore its scalability. For instance, co-expression of a dominant negative peptide of the *MLH1* gene has been shown to boost prime editing efficiency.[4] In addition, it has been shown that a limiting factor for prime editing is the degradation of pegRNAs; this led to the development of engineered pegRNAs (epegRNAs) which contain a stabilizing structured RNA motif.[5]

The eventual goal, in general, for most MAVEs is to catalog the functional effect of every possible missense variant in a particular gene of interest. This provides saturation-level data on functional consequences not only for variants in ClinVar, but also most possible new variants that will emerge as we continue to sequence more individuals. However, saturation-level MAVEs are expensive and require significant expertise. Herein we present curated loci prime editing (cliPE), a method that compromises saturation level data to focus specifically on VUS resolution of variants present in ClinVar. CliPE is a modular protocol adapted from SPE that is relatively inexpensive and has a low barrier to entry.[2] The data generated by cliPE has utility for variant resolution on its own, as most of the variants tested are reported VUS. We posit that cliPE is further well-suited to proof-of-concept or feasibility studies which can be used in proposals seeking funding for a saturation-level MAVE. We provide a series of internal positive controls that can be used to gauge the success of the prime editing component (**Supplemental Tables 3-4**). We also developed a user-friendly cliPE pegRNA Designer companion Shiny app (https://design.clipe-mave.org/) that can be used to import variant data from gnomAD and ClinVar and design the library of epegRNAs ready for order from the users oligonucleotide synthesis company of choice. It is important to note that the selection portion of cliPE (Module 5), like any MAVE approach, will be context-dependent with the gene of interest largely influencing the correct selection paradigm. While this protocol was optimized with the HAP1 cell line, high transferability to other cell lines is expected, though optimization may be required. We have validated cliPE with a *TSC2* MAVE and have reported that data elsewhere.[6]

# Before you begin

*Workflow overview*

The workflow involves 7 modules: (**Graphical Abstract**). Module 1 consists of inputting information into a pre-programmed Shiny app and downloading output design files including candidate epegRNA libraries, archetypal epegRNAs, and nicking gRNAs. Module 2 involves cloning archetypal epegRNAs into an expression plasmid, transfection into HEK 293T cells, and sequencing amplicons to estimate editing efficiency. Module 3 consists of cloning the epegRNA libraries based on the top-performing archetypal epegRNAs from Module 2 as well as cloning the matching nicking gRNAs. Modules 4 and 5 involve co-transfection of cells to induce genome editing followed by selecting cells based on the function of the gene of interest. Finally, in Modules 6 and 7, selected and unselected cells are sequenced using high-depth amplicon sequencing followed by enrichment analysis to identify variants enriched in selected cell pools relative to unselected control cell pools.

# Key resources table

| **REAGENT or RESOURCE** | **SOURCE** | **IDENTIFIER** |
|---|---|---|
| Chemicals, peptides, and recombinant proteins | | |
| Luria broth (LB) | Fisher | #L2542 |
| Ampicillin (amp) | Sigma | #A9518 |
| LB + amp plates (agar) | Make with Addgene protocol | https://www.addgene.org/protocols/pouring-lb-agar-plates/ |
| LB + amp broth | Dilute stock amp to 100 ug/mL in LB | n/a |
| Tryp-LE | Gibco | #25300062 |
| iProof PCR mastermix | Bio-Rad | #1725310 |
| Plasmid DNA extraction kit for minipreps | Qiagen | #27106 |
| Plasmid DNA extraction kit for midipreps | ZymoResearch | #D4200 |
| T4 DNA Ligase | NEB | #M0202S |
| GoldenGate enzyme mix | NEB | #E1601S |
| BsaI-HFv2 enzyme | NEB | #R3733S |
| 10X rCutSmart Buffer | NEB | #B6004 |
| BsmBI-v2 enzyme | NEB | #R0739S |
| NEBuffer r3.1 | NEB | #B6003 |
| Genomic DNA extraction kit | Invitrogen | #K1820-02 |
| Gel DNA recovery kit | Qiagen | #28104 |
| Chemically competent OneShot TOP10 | ThermoFisher | #C404003 |
| AmpureXP size selection beads | Beckman Coulter | #A63881 |
| TurboFectin 8.0 transfection reagent | OriGene | #TF81005 |
| Critical commercial assays | | |
| Qubit high-sensitivity dsDNA reagent | ThermoFisher | #Q32854 |
| Deposited data | | |
| cliPE Github Repository | This paper | https://github.com/calhoujd/calhoujd.github.io |
| Experimental models: Cell lines | | |
| Human: HEK 293T cells | ATCC | CRL-321 |
| Human: HAP1 cells | Horizon | #C669 |
| Oligonucleotides | | |
| See **Supplementary Tables** | This paper | n/a |
| Recombinant DNA | | |
| pCMV-PEmax-P2A-GFP | Chen et al.[4] | RRID:Addgene180020 |
| pEF1a-hMLH1dn | Chen et al.[4] | RRID:Addgene174824 |
| pU6-tevopreq1-GG-acceptor | Nelson et al.[5] | RRID:Addgene174038 |
| BPK1520 | Kleinstiver et al.[7] | RRID:Addgene65777 |
| pU6-Sp-pegRNA-RNF2_+5GtoT | Anzalone et al.[3] | RRID:Addgene_135957 |
| pU6-sp-sgRNA-RNF2_+41nick | Anzalone et al.[3] | RRID:Addgene_135958 |
| Software and algorithms | | |
| Jellyfish | Marcais et al.[8] | https://github.com/gmarcais/Jellyfish |
| cliPE companion Shiny apps | This paper | http://home.clipe-mave.org |
| Other | | |
| Neon electroporator | Invitrogen | #NEON1S |
| Magnet for AmpureXP bead size selection of next-generation sequencing libraries | Sergi Lab Supplies | #1005a |
| Qubit™ fluorometer | ThermoFisher | #Q33238 |
| Standard laboratory molecular biology equipment (gel electrophoresis, thermocycler, water bath, centrifuge, etc) | n/a | n/a |
| Standard tissue culture equipment (incubator, biosafety cabinet, light microscope) | n/a | n/a |
| Access to fluorescence-activated cell sorting (FACS) equipment at nearby core facility | n/a | n/a |

# Materials and equipment setup

**HEK 293T cell growth media**

| **Reagent** | **Final concentration** | **Amount** |
|---|---|---|
| DMEM (Gibco #11995-073) | 89% | 44.5 mL |
| FBS (R&D Systems; 50-152-7067) | 10% | 5 mL |
| PenStrep (Gibco #15140122; 100X stock) | 1% | 0.5 mL |
| **Total** | **n/a** | **50 mL** |

**HAP1 cell growth media**

| **Reagent** | **Final concentration** | **Amount** |
|---|---|---|
| IMDM (Gibco #12440061) | 89% | 44.5 mL |
| FBS (R&D Systems; 50-152-7067) | 10% | 5 mL |
| PenStrep (Gibco #15140122; 100X stock) | 1% | 0.5 mL |
| **Total** | **n/a** | **50 mL** |

# Step-by-step method details

## Module 1: Design archetypal epegRNAs, epegRNA libraries, and nicking gRNAs

**Timing: 1 h**

In Module 1, all of the prime editing designs will be generated with a companion pegRNA Designer Shiny app.

1. Download necessary ClinVar missense input file. This file will be used to incorporate important controls into your cliPE experiment, namely known benign or likely benign (BLB) and pathogenic or likely pathogenic (PLP) variants.
    a. Navigate to https://www.ncbi.nlm.nih.gov/clinvar/ and search for gene name.
    b. Toggle on the filter for 'Missense'
    c. Download as tsv.

2. Download necessary gnomAD file. This file will be used to incorporate important controls into your cliPE experiment, especially synonymous variants present in the general population.
    a. Navigate to https://gnomad.broadinstitute.org/ and search for gene name.
    b. Use the checkboxes appearing underneath the gnomAD variants section to filter for 'Missense/inframe indel' and 'Synonymous' variants
    c. Export variants to csv.

3. Generate designs with the cliPE pegRNA Designer companion Shiny app.

a. Navigate to the cliPE pegRNA Designer app (https://design.clipe-mave.org/) and input basic information (gene name, RefSeq transcript ID, etc.).
b. Upload ClinVar missense file and gnomAD missense file.
c. In the "Include additional variants in editing windows:" section, it is generally recommended to check all four boxes. If any of those variant classes aren't important for a particular use case, they can be excluded by leaving the box unchecked.
d. Adjust Additional Options section as needed.

**CRITICAL**: In the Additional Options section, adjust the gnomAD minimal allele count accordingly with your gene of interest to select for BLB (negative control) alleles. For genes with autosomal dominant genetic disorders, generally a relatively low allele count threshold of 3-5 is sufficient. However, in the case of autosomal recessive disorders, this threshold may need to be adjusted significantly. For example, the classic p.F508del variant in *CFTR* has an allele count of nearly 20,000 in gnomAD.

e. Download output files containing prime editing designs.

**CRITICAL**: It is recommended to run the cliPE pegRNA Designer app twice for each gene, once with the "Introduce missense VUS variants" Design Strategy option and once with the "Introduce missense PLP/BLB variants" Design Strategy option. The former prioritizes regions of the gene with the highest density of VUS to maximize the value of each epegRNA library. The latter introduces the maximal number of control ClinVar BLB and PLP missense variants. Please see below in 3f for an important discussion of control variants necessary for a successful cliPE experiment.

f. Confirm regions contain sufficient control variants before proceeding

**CRITICAL**: It is important that the regions targeted also include additional classes of variants, or a truth set which will be key for validation of the MAVE during data analysis. Overall, for most MAVEs, two truth sets comprised of positive and negative controls are used to assess assay validity, which will be referred to as the assay validation truth set and the clinical truth set (see **Table 1**). The assay validation truth set consists of (1) synonymous and missense variants found in the general population in databases such as gnomAD (negative controls) and (2) premature truncation codon (PTC) variants (positive controls). It is recommended to include at least 20 synonymous and 20 PTC assay validation truth set variants in a cliPE experiment. The clinical truth set similarly consists of negative and positive controls present in the ClinVar database. The negative controls in the clinical truth set are BLB variants and the positive controls are PLP missense variants. It is recommended to include at least 25-30 clinical truth set variants in a cliPE experiment. Further discussion of this topic is available within the "Designing initial set of epegRNA architectures to screen, epegRNA libraries, and nicking gRNAs" subsection of the cliPE homepage (https://home.clipe-mave.org/).

g. Order oligonucleotides from preferred vendor for each archetypal epegRNA (top and bottom strand for each spacer, top and bottom strand of each extension), the pegRNA scaffold (this needs to be phosphorylated; sequence provided in **Supplemental Table 2**), and primers to screen for editing efficiency. Generally, one primer pair is needed to screen for editing for each archetypal epegRNA. However, if different regions of the same exon are targeted, a single pair of primers may be used to screen editing for more than one epegRNA.

**NOTE**: It is recommended to screen a minimum of 12 archetypal epegRNAs which will produce on average 3-6 epegRNA libraries which will be edit with high enough efficiency for the cliPE workflow. It may be desirable to screen more than 12 archetypal epegRNAs upfront to increase the probability of attaining enough epegRNA designs to proceed with library cloning. Further discussion of this topic is available within the "Designing initial set of epegRNA architectures to screen, epegRNA libraries, and nicking gRNAs" subsection of the cliPE homepage (https://home.clipe-mave.org/).

## Module 2: Screen archetypal epegRNAs representative of each candidate library

**Timing: 3-4 weeks**

In Module 2, archetypal epegRNAs which are representative of candidate epegRNA libraries will be cloned and tested for prime editing efficiency.

4. Cloning individual prime editing constructs.
    a. Set up the restriction digest for the destination vector: (1) 1 ug pU6-tevopreq1-GG-acceptor plasmid, (2) 5 uL 10X rCutSmart Buffer, (3) 1 uL BsaI-HFv2 enzyme, (4) X uL DNA grade water to a final reaction volume of 50 uL.
    **NOTE:** It may be advisable to scale this reaction up to generate sufficient digested plasmid. Each GoldenGate reaction here requires 60 ng of digested vector. Downstream cloning if epegRNAs in Module 3 require 50 ng of digested vector per epegRNA library. For example, cloning 10 archetypal epegRNAs and 5 epegRNA libraries will require a total of 850 ng of digested vector.
    b. Incubate for a minimum of 3-4 hours at 37°C followed by inactivation of enzymatic activity at 80°C for 20 min.
    **CRITICAL:** While many vendors recommend short (15 min) incubation times for restriction enzymes, our experience is that longer restriction digests lead to improved cloning efficiency.
    c. Gel purify the vector.
    **NOTE:** Reserve any leftover digested vector at -20°C for future ligations, including cloning of epegRNA libraries in Module 3.
    d. Pre-anneal components for GoldenGate cloning: (1) spacer duplex, (2) epegRNA extension duplex, and (3) phosphorylated scaffold. For each, add 1 uL of each oligo (100 uM stock) to 23 uL of DNA grade water. Then, heat on a thermocycler to 95°C, incubate for 3 min, and then cool slowly to 25°C (ramp speed: 5°C per min). Finally, add 75 uL to each tube to dilute to proper concentration for GoldenGate cloning.
    e. Set up GoldenGate cloning reaction: (1) 2 uL digested vector (30 ng/uL), (2) 2 uL annealed spacer oligos, (3) 2 uL annealed pegRNA extension oligos, (4) 2 uL annealed phosphorylated scaffold, (5) 1 uL GoldenGate enzyme mix, (6) 2 uL 10X T4 DNA ligase buffer, and (7) 9 uL DNA grade water. Incubate on thermocycler at 37°C for 1 hour followed by 60°C for 5 min.
    f. Transform bacteria
        i. Thaw bacteria on ice for up to 20 min.
        ii. Add 1 uL of the GoldenGate reaction to 25-50 uL of chemically competent bacteria. Flick gently to mix.

iii. Incubate on ice up to 20 min.
iv. Perform heat shock by rapidly transferring to 42°C water bath for 30 sec followed by rapid transfer back to ice for at least 2 min.
v. Add 300 uL SOC media and incubate at 37°C for 30-60 min with shaking at 200-250 rpm.
vi. Pre-warm 1-2 LB+Amp plates per cloning reaction to room temperature.
vii. Plate bacteria onto pre-warmed plates.
**NOTE:** It is recommended to plate low (25-50 uL) and high (100-200 uL) volumes of each cloning reaction to increase likelihood of easily picking single colonies across a range of cloning efficiencies.
viii. Incubate plates overnight (or about 16 hrs) at 37°C.
ix. Pick 3-4 colonies per archetypal epegRNA and sequence by Sanger to QC correct cloning of each construct.

5. Prime editing in HEK 293T cells
   a. Design and order primers for sequencing of regions targeted for genome editing.
   **NOTE:** It is important to ensure that the primers do not bind too close to the site of editing, particularly for primer designs for Sanger sequencing, due to the extra noise in the first 25-35 bases of sequencing data. An amplicon size of 500-700 bp is optimal for Sanger sequencing, while amplicons of 800-1200 bp are optimal for LR sequencing. We recommend using software such as Primer3 (https://primer3.ut.ee/) to aid with primer design.
   b. Seed 100,000 cells 16-24 hours before transfection in 24-well plates.
   c. Transiently co-transfect cells using either TurboFectin 8.0 or another suitable transfection reagent (see **Table 2** for amounts of each vector).
   d. 24-48 hours post-transfection, sort GFP+ cells at a flow cytometry core facility. It is recommended to always plate an extra well of control untransfected cells for the purpose of setting the gate for GFP+ cells and accounting for autofluorescence. It is recommended to sort at least 10,000-50,000 cells per condition.
   e. After sorting, re-plate GFP+ cells under standard culture conditions for at least 48 hours.
   f. Collect cell pellets by either scraping or trypsinization with Tryp-LE and centrifugation in 1.5 mL Eppendorf tubes at 300xg for 5 min at RT.
   **NOTE:** Pellets can be either used immediately for genomic DNA (gDNA) extraction or frozen at –20°C prior to gDNA extraction.
   g. Extract gDNA using standard gDNA miniprep kit.
   h. Perform screening by PCR.
      i. Setup the PCR reaction. After aliquoting 22 uL of mastermix in each tube of PCR strips, add 20-60 ng template gDNA (3 uL of 6.67-20 ng/uL). It is recommended for each primer pair to include an untransfected control and a no template control in addition to the cells co-transfected for prime editing.
      **OPTIONAL:** Run a small aliquot (~5 uL) on a 1-2% agarose gel to confirm amplification.

**PCR Reaction Mastermix**

| Reagent | Amount |
|---|---|
| 2X iProof Mastermix | 12.5 uL |
| forward primer (10 uM) | 1.25 uL |
| reverse primer (10 uM) | 1.25 uL |
| $ddH_2O$ | To 22 uL total volume |

**PCR Cycling Conditions**

| Steps | Temperature | Time | Cycles |
|---|---|---|---|
| Initial Denaturation | 98 °C | 3 min | 1 |
| Denaturation | 98 °C | 20 sec | 30-35 cycles |
| Annealing | 62 °C | 20 sec | |
| Extension | 72 °C | 30 sec | |
| Final extension | 72 °C | 7 min | 1 |
| Hold | 4 °C | forever | |

ii. PCR cleanup amplicons to prepare for sequencing.
**NOTE:** Quick cleanup with a column-based kit is usually sufficient. If primers produce multiple amplicons which interfere with downstream sequencing, it may be necessary to perform gel electrophoresis and gel purify the specific amplicon for sequencing.

iii. Submit amplicons for either Sanger or long-read (LR) sequencing (vendor such as Plasmidsaurus).
**NOTE:** Carefully follow recommendations provided by the sequencing vendor to properly submit samples.

i. Select epegRNAs with sufficient editing (>15% either confirmed by LR or estimated by chromatogram peak height) for downstream cliPE experiment.

j. Confirm regions contain sufficient control variants before proceeding

**CRITICAL**: Once the low-performing epegRNAs are filtered out, it is important to reassess the presence of control set variants as discussed above in 3d-f. If the libraries do not contain enough control set variants, it is recommended to revisit the pegRNA Designer app, design a second round of prime editing constructs, and do one more round of archetypal epegRNA screening to ensure the final dataset will contain a minimal number of control variants.

## Module 3: Clone nicking gRNAs and prime editing libraries

**Timing: 4-5 weeks**

In Module 3, epegRNA libraries and nicking gRNA vectors will be cloned and validated.

6. Clone nicking gRNA constructs.
   a. Order oligonucleotides with appropriate sticky ends.
   **NOTE:** These are provided in the output from the cliPE pegRNA Designer Shiny app.
   **TROUBLESHOOTING:** Cloning of nicking gRNAs into BPK1520 follows a similar principle as standard px458/px459 cloning. See problem 2 for how to avoid a common pitfall in gRNA cloning.
   b. Linearize vector BPK1520: (1) 1 ug BPK1520, (2) 2 uL 10X NEBuffer r3.1, (3) 1 uL BsmBI-HF enzyme, (4) 16 uL DNA grade water. Incubate for a minimum of 3-4 hours at 55°C followed by inactivation of enzymatic activity at 80°C for 20 min. Purify by gel or column purification. It is recommended to examine a small aliquot by gel electrophoresis (0.8-1% agarose) to confirm complete digestion of vector. It is further recommended to scale this reaction up further in anticipation of additional nicking gRNA cloning reactions. Leftover linearized vector can be stored at -20°C when not in use.

c. Setup the phosphorylation and annealing reaction: (1) 1 uL oligo 1 (100uM), (2) 1 uL oligo 2 (100uM), (3) 1 uL 10X T4 Ligation Buffer (NEB) (4) 6.5 uL DNA grade water, (5) 0.5 uL T4 PNK (NEB). Please note it is important to use T4 Ligation Buffer rather than PNK Buffer. Incubate in a thermocycler at 37°C for 30 min to anneal oligos. Then, heat to 95°C and then cool slowly to 25°C (ramp speed: 5°C per min).
d. Set up the ligation reaction: (1) 50 ng BbsI-linearized expression vector BPK1520, (2) 1 uL phosphorylated and annealed oligo duplex (**1:200 dilution**), (3) 1 uL 10X T4 DNA ligase buffer (NEB), (4) X uL ddH2O (to a total of 10 uL), (5) 1 uL T4 DNA Ligase (NEB). Incubate for 15 min at RT.
e. Transform bacteria.
    i. Thaw bacteria on ice for up to 20 min.
    ii. Add 1-2 uL of ligated plasmid to 25-50 uL of chemically competent bacteria. Flick gently to mix.
    iii. Incubate on ice up to 20 min.
    iv. Perform heat shock by rapidly transferring to 42°C water bath for 30 sec followed by rapid transfer back to ice for at least 2 min.
    v. Add 300 uL SOC media and incubate at 37°C for 30-60 min with shaking at 200-250 rpm.
    vi. Pre-warm 1-2 LB+Amp plates per cloning reaction to room temperature.
    vii. Plate bacteria onto pre-warmed plates.
    **NOTE:** It is recommended to plate low (25-50 uL) and high (100-200 uL) volumes of each cloning reaction to increase likelihood of easily picking single colonies across a range of cloning efficiencies.
    viii. Incubate plates overnight (or about 16 hrs) at 37°C.
    ix. Pick 3-4 colonies per nicking gRNA and sequence by Sanger to QC correct cloning of each construct.

7. Cloning epegRNA libraries.
    a. Order single-stranded DNA oligo pools and primers to generate double-stranded oligo pool and append BsaI restriction enzyme recognition sites.
    **NOTE:** These are provided in the output from the cliPE pegRNA Designer Shiny app.
    b. Resuspend pools of single-stranded DNA oligos (IDT oPools) encoding epegRNAs with 50 uL of DNA grade water or 1X TE. Make a working stock by diluting to 20 ng/uL.
    **NOTE:** For IDT oPools, the stock concentration will be equivalent to the number of oligos in the pool in uM (i.e., a 50x oligo pool will be 50 uM).
    c. Set up PCR reactions to generate double-stranded oligo pool and append BsaI restriction enzyme recognition sites (10 cycles).

**PCR Reaction Mastermix**

| Reagent | Amount |
|---|---|
| Oligo pool (20 ng/uL) | 3 |
| 2X iProof Mastermix | 12.5 uL |
| forward primer (10 uM) | 1.25 uL |
| reverse primer (10 uM) | 1.25 uL |
| $ddH_2O$ | 7 uL total volume |

**PCR Cycling Conditions**

| Steps | Temperature | Time | Cycles |
|---|---|---|---|
| Initial Denaturation | 98 °C | 2 min | 1 |
| Denaturation | 98 °C | 30 sec | 10 cycles |
| Annealing | 60 °C | 10 sec | |
| Extension | 72 °C | 30 sec | |
| Final extension | 72 °C | 7 min | 1 |
| Hold | 4 °C | forever | |

**NOTE:** It may be necessary to adjust the annealing temperature to optimize for some primer pairs.

d. Use a PCR cleanup kit such as Qiagen's QIAquick PCR and gel cleanup kit (#28104) to quickly prepare the amplified DNA for restriction digest. Elute in 30 uL of elution buffer.

e. Set up the restriction digest to generate sticky ends for cloning: (1) 30 uL PCR-amplified oligo pool, (2) 5 uL 10X NEB rCutSmart Buffer, (3) 1 uL BsaI-HFv2 enzyme (NEB; #R3733S), (4) 14 uL DNA grade water. Incubate for at least 3-4 hours at 37°C followed by inactivation of enzymatic activity at 80°C for 20 min. Purify by column purification.

f. If necessary, set up the restriction digest for the destination vector as outlined above in step 4a-c.

g. Set up the ligation reaction: (1) 37.5 ng digested and purified epegRNA pool, (2) 50 ng digested and purified destination vector, (3) 2 uL 10X T4 DNA ligase buffer (NEB), (4) X uL DNA grade water (to a total of 20 uL), (5) 1 uL T4 DNA Ligase. Incubate for 15 min at RT.

h. Transform 1-2 uL ligated plasmid pool into competent bacteria and plate onto LB+Amp plates.

i. Culture plates overnight (about 16 hr). Carefully add 1 mL of LB and scrape colonies into a 1.5 mL tube and pellet (8000xg for 3 min at RT). Use a midiprep kit to extract plasmid DNA from the pooled colonies.

j. Determine library concentrations with the Qubit high-sensitivity dsDNA reagent.

k. Perform a first round of QC using a plasmid LR sequencing service such as Plasmidsaurus. Determine whether epegRNA architecture is correct and if different variants are present as expected.

**OPTIONAL:** Perform a second round of QC using targeted amplicon sequencing of each epegRNA to estimate the frequency of each epegRNA in each plasmid pool. Long-read (LR) sequencing services like Plasmidsaurus are useful for initial QC of an epegRNA library, but it is difficult to accurately estimate the frequency of each epegRNA within the pool with the limited number of reads you routinely receive. Illumina targeted amplicon sequencing will give sufficient read depth to quantify the frequency of each epegRNA. The following primers are recommended for submitting as part of a large, multiplexed batch of ampSeq libraries with either single or double-barcoding (Illumina adaptors are underlined):

tevopreqPCR1_ampSeqF: (5'-
TCGTCGGCAGCGTCAGATGTGTATAAGAGACAGATATATCTTGTGGAAAGGACGAAAC-3')
tevopreqPCR1_ampSeqR: (5'-
GTCTCGTGGGCTCGGAGATGTGTATAAGAGACAGTACCTCGAGCGGCCCA-3')

Please note, if sending a small batch of sequencing to a service such as GENEWIZ from Azenta Amplicon-EZ or the MGH CCIB DNA Core's Complete Amplicon Sequencing, it is advised to adjust the primers to be compatible. Please carefully read the sample submission instructions for the particular service.

## Module 4: *Prime edit haploid HAP1s*

**Timing: 1 h**

In Module 4, haploid cells are co-transfected to generate cell pools with each cell containing a single variant in the gene of interest.

8. Generate cell library containing individual variants with prime editing.
   a. Culture enough dishes of HAP1 cells for the number of transfections needed (1 million cells per electroporation; yield is approximately 10 million cells per standard 10 cm culture dish).
   b. Transiently co-transfect cells using the Invitrogen Neon electroporator using the 100 uL kit (#MPK10096). Please see **Table 3** for suggested amounts of each vector.
   c. 24 hours post-transfection, sort GFP+ cells at a flow cytometry core facility. It is recommended to sort 50,000-200,000 cells per co-transfected cliPE library.
   d. After sorting, re-plate GFP+ cells under standard culture conditions for at least 48 hours.
   **OPTIONAL:** Perform a QC check by performing amplicon sequencing of cells to confirm genome editing of target region. This data can be redundant with data from unselected cells in Module 5, but it may be useful to validate the cell library prior to preceding with selection.

## Module 5: Select cells

**Timing: Variable.**

In Module 5, cell pools generated are selected by a chosen method which will enrich for cells containing a certain class of variants, such as those containing loss-of-function of the gene of interest. This Module is highly context-dependent and as such, it is not possible to provide an single protocol that will work for any gene of interest; instead, we have provided below a number of resources that can guide in selecting an appropriate selection method.

9. Select cells by growth assay, cell sorting, or other method.
   **CRITICAL:** There are some generalizable methods that are worth considering first as a potential selection strategy for a MAVE. If the gene of interest is an essential gene in HAP1s or another cell type, this can be exploited to perform a MAVE using genome editing followed by culturing the cells for different amounts of time.[9] After a certain amount of time in culture, LOF alleles will be depleted as cells expressing these variants fail to thrive. This is a very popular selection strategy and there are many examples in the literature to use to guide your experimental design.[10,11] If the gene is not an essential gene, it still might be possible to find a cell line where growth is partially dependent on the gene of interest; DepMap is a useful resource with genome-wide data on many cell lines using both gene knockout and mRNA knockdown approaches (https://depmap.org/portal/). If pathogenicity of the gene of interest correlates with reduced protein abundance, VAMPseq is a powerful method to identify pathogenic variants, as demonstrated in previous studies on PTEN and others.[12] Additional generalizable methods include response to drug treatment, single cell transcriptomics, cell surface trafficking of membrane proteins, and cell morphology.[13-18] If none of these more generalizable methods are a fit, there are certainly other gene-specific alternatives you can employ. Sorting cells based on cell signaling activity enabled us to distinguish pathogenic *TSC2* missense variants.[6] The BioGrid ORCS database (https://orcs.thebiogrid.org/) of high-throughput CRISPR screens can identify potential selection approaches to facilitate a MAVE study. We also recommend reading through reviews such as Starita L et al. and Tabet D et al. which provide excellent guidance on selection strategies for high-throughput functional studies.[19,20]

**NOTE:** For new assays, it is recommended to perform pilot experiments with a single variant or small number of variants, such as individual epegRNAs designed to generate control variants such as missense BLB and PLP variants, synonymous variants present in the general population, or loss-of-function variants produced by PTC codons.

## Module 6: Sequence regions of genes targeted for genome editing

**Timing: Variable**

In Module 6, next-generation targeted amplicon sequencing of each region selected for prime editing is performed in both control unselected cells as well as cells which have undergone selection. For example, if a growth assay was performed, libraries might be constructed at an early and a late timepoint after prime editing.

10. Prepare amplicon sequencing library (or libraries).
    **CRITICAL:** Use step 10 for smaller batch preparation when sequencing one or a small number of libraries.
    a. Design primers to amplify the region spanning the reverse transcription (RT) template of each epegRNA architecture. See above for advice on designing these primers.
       **NOTE:** It is necessary to design primers to amplify each locus targeted for genome editing. It is critical to constrain the total amplicon size to be less than 250 bp to maximize read depth of the target region. Also, as above, it is important to ensure the primers do not overlap the region of interest to detect any small insertions and deletions (indels). If sequencing with a vendor such as MGH CCIB DNA Core's Complete Amplicon Sequencing service or GENEWIZ from Azenta Amplicon-EZ, it is important to review the sample submission guidelines specific to the respective service.
    b. Extract gDNA from pellets of cells collected under selection conditions. Make sure to also extract gDNA from cell pellets without selection as a control.

c. Set up PCR reaction to generate amplicons for targeted amplicon sequencing. After aliquoting 22 uL of mastermix in each tube of PCR strips, add 20-60 ng template gDNA (3 uL of 6.67-20 ng/uL).

**PCR Reaction Mastermix**

| Reagent | Amount |
|---|---|
| 2X iProof Mastermix | 12.5 uL |
| ampSeq_FwdPrimer (10 uM) | 1.25 uL |
| ampSeq_RevPrimer (10 uM) | 1.25 uL |
| $ddH_2O$ | To 22 uL total volume |

**PCR Cycling Conditions**

| Steps | Temperature | Time | Cycles |
|---|---|---|---|
| Initial Denaturation | 98 °C | 3 min | 1 |
| Denaturation | 98 °C | 20 sec | 30-35 cycles |
| Annealing | 60 °C | 20 sec | |
| Extension | 72 °C | 30 sec | |
| Final extension | 72 °C | 7 min | 1 |
| Hold | 4 °C | forever | |

d. Use size selection beads to remove primer dimer and prepare libraries for sequencing. AmpureXP size selection beads are routinely used to remove primer dimer and prepare NGS libraries for either subsequent PCR or for Illumina sequencing.
   i. Allow aliquots of AmpureXP beads to sit at room temperature (RT) for at least 30 min prior to use.
   ii. Add 25 uL of DNA grade water to each 25 uL PCR reaction to bring the total volume to 50 uL. Add an equal volume (50 uL) of AmpureXP beads to bring the total volume to 100 uL and mix well by pipetting.
   iii. After a 10 min incubation at RT, place on magnet for 5 min.
   iv. Remove and discard supernatant, followed by two washes with 100-200 uL of 70% ethanol.
   v. Allow beads to air dry for about 5 min.
   vi. Add 21 uL of 1X TE and incubate for 5 min prior to placing tubes back on magnet.
   vii. After 5 min of separation, pipette 20 uL of eluted DNA into a fresh tube.

e. Use Qubit to quantify DNA concentration for each pool of amplicons, dilute to appropriate concentration, and submit for Illumina short-read sequencing.

11. Prepare amplicon sequencing libraries.
    **CRITICAL:** Use step 11 for larger batches of many libraries sequenced in multiplex.
    a. Design primers to amplify the region spanning the RT template of each epegRNA architecture. See step 10A for specific design details. Append Illumina adaptors prior to ordering. If necessary, order primers for the second PCR (barcoding PCR) reaction described below (see **Supplemental Table 1** for primer sequences to order from IDT or preferred vendor).

**NOTE:** To sequence in multiplex at a core facility or outside vendor, it is necessary to append the appropriate Illumina adaptors to enable barcoding:

Primer 1: adapter + forward target primer (5'-TCGTCGGCAGCGTCAGATGTGTATAAGAGACAG -forward_primer-3').

Primer 2: adapter + reverse target primer (5'-GTCTCGTGGGCTCGGAGATGTGTATAAGAGACAG -reverse_primer-3').

b. Extract gDNA from unselected and selected cell pellets.
c. Set up PCR reaction to generate initial amplicons for targeted amplicon sequencing. After aliquoting 22 uL of mastermix in each tube of PCR strips, add 20-60 ng template gDNA (3 uL of 6.67-20 ng/uL).

**PCR Reaction Mastermix**

| Reagent | Amount |
|---|---|
| DNA template | 60 ng |
| 2X iProof Mastermix | 12.5 uL |
| ampSeq_FwdPrimer (10 uM) | 1.25 uL |
| ampSeq_RevPrimer (10 uM) | 1.25 uL |
| $ddH_2O$ | To 25 uL total volume |

**PCR Cycling Conditions**

| Steps | Temperature | Time | Cycles |
|---|---|---|---|
| Initial Denaturation | 98 °C | 3 min | 1 |
| Denaturation | 98 °C | 20 sec | 30-35 cycles |
| Annealing | 60-62 °C | 20 sec | |
| Extension | 72 °C | 30 sec | |
| Final extension | 72 °C | 7 min | 1 |
| Hold | 4 °C | forever | |

d. Use size selection beads to remove primer dimer and prepare libraries for barcoding PCR.
   i. Allow aliquots of AmpureXP beads to sit at room temperature (RT) for at least 30 min prior to use.
   ii. Add 25 uL of DNA grade water to each 25 uL PCR reaction to bring the total volume to 50 uL. Add an equal volume (50 uL) of AmpureXP beads to bring the total volume to 100 uL and mix well by pipetting.
   iii. After a 10 min incubation at RT, place on magnet for 5 min.
   iv. Remove and discard supernatant, followed by two washes with 100-200 uL of 70% ethanol.
   v. Allow beads to air dry for about 5 min.
   vi. Add 21 uL of 1X TE and incubate for 5 min prior to placing tubes back on magnet.

vii. After 5 min of separation, pipette 20 uL of eluted DNA into a fresh tube.

**OPTIONAL:** Perform a qPCR reaction to optimize the number of cycles for barcoding PCR. 2 uL of PCR1 product (1:20 dilution) are added directly to each well. Select a number of cycles early in the linear phase. Include a step to read SybrGreen fluorescence after each round of PCR. Optimal cycle numbers typically vary between 7 and 15. If omitting this step, it is recommended to use a standard 12 cycles which is typically sufficient for barcoding amplicon sequencing libraries.

**PCR Reaction Mastermix**

| Reagent | Amount |
|---|---|
| 2X iProof Mastermix | 12.5 uL |
| ampSeqPCR2univFwdPrimer (2.5 uM) | 3 uL |
| ampSeqPCR2univRevPrimer (2.5 uM) | 3 uL |
| SybrGreen (100X) | 0.25 uL |
| $ddH_2O$ | To 23 uL total volume |

**PCR Cycling Conditions**

| Steps | Temperature | Time | Cycles |
|---|---|---|---|
| Initial Denaturation | 98 °C | 2 min | 1 |
| Denaturation | 98 °C | 30 sec | 40 cycles |
| Annealing | 60 °C | 10 sec | |
| Extension | 72 °C | 30 sec | |
| Final extension | 72 °C | 7 min | 1 |
| Hold | 4 °C | forever | |

e. Set up PCR2 barcoding reaction.

**CRITICAL:** It is important that each sample has a unique combination of i5 and i7 indexes for proper demultiplexing. It is imperative to make a spreadsheet of which samples are receiving which barcodes both for correct pipetting as well as downstream demultiplexing. For single index multiplexing, up to 24 amplicon sequencing libraries can be pooled together. Select a constant i5 index (one of P5_[…] in **Supplemental Table 1**) and up to 24 of the i7 indeces (P7_[…] in **Supplemental Table 1**). For double barcoded libraries, up to 384 samples can be pooled together. Separate pcr mastermixes are made for each unique i5 index. 22 uL of mastermix are than aliquoted into PCR strip tubes. 2 uL of PCR1 product (1:20 dilution) are added directly to each tube. 3 uL of i7 index primer (2.5 uM) are added directly to each tube.

**PCR Reaction Mastermix**

| Reagent | Amount |
|---|---|
| 2X iProof Mastermix | 12.5 uL |
| i5 index primer (2.5 uM) | 3 uL |
| $ddH_2O$ | To 20 uL total volume |

**PCR Cycling Conditions**

| Steps | Temperature | Time | Cycles |
|---|---|---|---|
| Initial Denaturation | 98 °C | 2 min | 1 |
| Denaturation | 98 °C | 30 sec | 12 cycles or number determined during optional qPCR |
| Annealing | 60 °C | 10 sec | |
| Extension | 72 °C | 30 sec | |
| Final extension | 72 °C | 7 min | 1 |
| Hold | 4 °C | forever | |

f. Pool 5-10 uL of each library and use AmpureXP bead purification to clean up amplicon pool prior to sequencing.
   i. Allow aliquots of AmpureXP beads to sit at room temperature (RT) for at least 30 min prior to use.
   ii. Add 0.9X volume of AmpureXP beads to pooled library. For example, if pooling 10 libraries and using 10 uL per library, add 90 uL of AmpureXP beads. Mix well by pipetting.
   iii. After a 10 min incubation at RT, place on magnet for 5 min.
   iv. Remove and discard supernatant, followed by two washes with 100-500 uL 70% ethanol. Use enough volume of 70% ethanol to cover beads.
   v. Allow beads to air dry for about 5 min.
   vi. Add 21 uL of 1X TE and incubate for 5 min prior to placing tubes back on magnet.
   vii. After 5 min of separation, pipette 20 uL of eluted DNA into a fresh tube.

g. Use Qubit to quantify DNA concentration for final amplicon pool, dilute to appropriate concentration, and submit for Illumina short-read sequencing.
   **NOTE:** Follow the provided sample submission instructions for custom short-read Illumina libraries from sequencing vendor. It is critical to request 2x 150 bp (i.e., paired-end) sequencing on an appropriate platform to get a minimum of 200,000 reads per individual amplicon sequencing library in the pool. No custom sequencing primers are necessary, although you will need to provide the indices used for barcoding (see **Supplementary Table 1**). It is also critical to let the sequencing facility know that these libraries likely qualify as low sequencing diversity libraries and may require higher than normal spike-in of PhiX, unless they will be multiplexed with other libraries from users with diverse libraries. Please see the cliPE GitHub repository (https://github.com/calhoujd/calhoujd.github.io) for example sample sheets, library kits, and demultiplexing scripts using bcl2fastq.

## Module 7: Analysis of data using random effects modeling

**Timing: 1-2 days**

In Module 7, allele frequencies are computed from unselected and selected cell pools. The resulting enrichment score, if successful, will distinguish pathogenic from benign variants in the gene of interest.

12. Perform k-mer counting and enrichment analysis.

a. If not installed previously, install the Jellyfish k-mer counting software (https://github.com/gmarcais/Jellyfish).[8]

b. Unzip read1 fastqs:

```
>gunzip read1.fastq.gz
```

**NOTE:** It is recommended to write a script which combines this and the following several steps to run with one command using either sh, msub, or sbatch, depending on whether you are running this locally or on a high-performance compute cluster. Please see the cliPE GitHub repository (https://github.com/calhoujd/calhoujd.github.io) for an example script written for sbatch for Slurm scheduler.

c. Run the Jellyfish *bc* command on read1 fastqs:

```
>jellyfish bc -m <k-mer length> -s 1G -t <threads> -o
read1.bc -C read1.fastq
```

d. Run Jellyfish *count* command:

```
>jellyfish count -m <k-mer length> -s 1G -t <threads> -o
read1_MERcounts.jf --bc read1.bc --if kmerLibrary.fasta -C
read1.fastq
```

**NOTE:** The fasta library is generated previously by the Shiny cliPE pegRNA Designer app.

e. Run the Jellyfish *dump* command:

```
>jellyfish dump read1_MERcounts.jf >
mer_counts_dump_read1.fa
```

f. Repeat steps b through e for read 2 for the same sample, and then repeat this process for each pair of fastq files. It is important to note this can be streamlined using standard file naming conventions and running all jobs simultaneously using programmatic job submission such as job arrays.

g. Run the cliPEr_app1_fasta2csv app (https://calhoujd12.shinyapps.io/cliPEr_app1_fasta2csv/) to convert fasta file to csv file. Repeat for all read1 and read2 fastqs.

h. Annotate the k-mer count file with the variant name. Upload the output from cliPEr_app1_fasta2csv as well as the dictionary generated previously in module 1 to the Shiny cliPEr_app2_kmers2variants app (https://calhoujd12.shinyapps.io/cliPEr_app2_kmers2variants/). If the Jellyfish output contains the reverse complement k-mer relative to the dictionary generated by cliPEpy_1 above, there is an option to reverse complement within the cliPEr_app2_kmers2variants Shiny app to ensure successful annotation.

i. Generate a final csv for input into the third R-based Shiny app. Each epegRNA architecture will have a separate file. Column1 will be the name of each variant. Column2 will be ‘counts_control_rep1’, the counts for the variant in cells without selection. Column3 will be ‘counts_selected_rep1’, the counts for the variant in cells after selection. Additional replicates will each be represented by additional pairs of columns. It is recommended to include 3-4 biological replicates for each epegRNA library.

j. Upload csv to the cliPEr_app3_random_effects_modeling companion Shiny app (https://calhoujd12.shinyapps.io/cliPEr_app3_random_effects_modeling/). You will also need to specify the number of biological replicates. For each variant, the allele frequency will be determined relative to the depth of sequencing. A random effects model is generated using the mclust R package. Random effects modeling has been recommended previously for MAVE data analysis with the Enrich2 statistical framework.[21] After the analysis is complete, download the output csv file which will contain two additional columns to the original input: (1) the beta, or functional enrichment score, and (2) the standard error for each variant.

**NOTE:** Alternatively, other tools are available which can perform random effects modeling for MAVE data, such as Enrich2 (https://github.com/FowlerLab/Enrich2) or CountESS (https://github.com/CountESS-Project/CountESS).[21]

k. It is recommended to filter out variants which have a low count in unselected cells due to either low editing efficiency or low abundance of certain epegRNAs in the final epegRNA pool. In our experience, filtering out variants in unselected cells below 0.1% allele frequency is necessary as variants below this cutoff tend to have enrichment scores outside of the ranges for variants in the same class, as well as high variability between replicates. A further discussion of this phenomenon can be found in Rubin et al.[21] This threshold may need to be adjusted somewhat for each gene:selection pair and can be done by assessing the enrichment scores and SE for internal assay validation variants with varying cutoff thresholds. Optionally, it may be of interest to normalize functional scores across epegRNA libraries, similar to the normalization in Buckley et al.[22] It is important at this step to critically assess whether the functional scores are performing as intended on included internal assay controls. Namely, are synonymous and PTC variants well-separated? How do the functional scores for ClinVar BLB variants compare to those of ClinVar PLP variants? At what level of evidence can this data be used in a variant classification framework? We recommend using guidelines outlined in Brnich et al. for calibrating cliPE MAVE data.[23]

## Expected outcomes

In Module 1 and using the pegRNA Designer Shiny app, users will be able to quickly generate cliPE designs for a gene of interest. In Module 2, archetypal epegRNAs are screened to filter out epegRNAs which do not produce robust genome editing. It is recommended to screen a minimum of 12 archetypal epegRNAs, as the dropout rate at this stage is estimated to be 50-75%. We provide an example archetypal epegRNA screen in **Figure 1**. As machine learning predictions of prime editing efficiency improve, it may become possible to reduce the dropout rate at this stage. Before proceeding to Module 3, we encourage users to confirm that the epegRNA libraries corresponding to the validated archetypal epegRNAs contain sufficient variants, including the different classes of truth set variants discussed above (with additional context provided on https://home.clipe-mave.org/). It may be necessary to screen an additional set of archetypal epegRNAs to fill in gaps in the truth set. The Shiny app provided in Module 1 allows prioritization for either VUS or truth set variants, which may be useful if the regions of the gene with highest VUS density do not contain sufficient truth set variants.

After completing cloning steps in Module 3, it is anticipated that users will obtain epegRNA libraries and nicking gRNAs and can proceed directly to co-transfection of haploid cells in Module 4. The optional QC step of targeted amplicon sequencing of epegRNA libraries is useful to check for potential variant dropout due to particular RT template oligos cloning inefficiently. Library composition is usually not a major confounding factor for cliPE as each library is relatively small (on the order of 40-60 unique epegRNAs). Optionally, the QC step outlined in Module 4 is the most useful step to check for variant dropout by amplicon sequencing of the target locus in GFP+, co-transfected cells. Based on our previous work using cliPE on the *TSC2* gene, it is anticipated that overall variant dropout rate will be ~33%. We provide an example of prime editing with an epegRNA library in **Figure 2**.

The selection portion of the cliPE workflow is highly context-dependent, and we anticipate cliPE will be compatible with a wide variety of selection paradigms. The most important consideration is to ensure that the selection method is compatible with collection of sufficient gDNA from selected and unselected cell pools. The yield of gDNA from selected cell pools needs to be compatible with PCR amplification of the target locus for downstream amplicon sequencing. This can be successful even with relatively low yields in the 0.2-1.0 ng/uL range.

We anticipate most individuals will utilize Module 6A mainly to QC library composition of epegRNA libraries and QC editing rate of epegRNA libraries prior to selection (Module 5). Once sufficient QC has been completed, we envision Module 6B will be the general use case for sequencing selected and unselected cell pools generated in Module 5. For a small number of libraries, the 6A protocol is efficient and relatively cost-efficient, but the cost does not scale competitively once multiple biological replicates are being generated across multiple epegRNA libraries. The 6B protocol allows for multiplexing hundreds of samples followed by sequencing a single pooled library on the Illumina platform. This is highly cost-efficient as only a minimum of 200,000 reads are recommended per sample. The analysis in Module 7 is streamlined and requires installation of only a single package, the Jellyfish software for k-mer counting. The Jellyfish outputs are then fed into a series of Shiny apps to generate enrichment scores by random effects modeling, similar to the Enrich2 software package.[21] The code underlying these Shiny apps are shared on the cliPE Github page and users are welcome to modify as necessary to suit the needs of their experiment.

*Estimating the cost of cliPE*

With a small number of epegRNA libraries (5-10), cliPE is anticipated to cost about $1300 per library. Among the biggest expenses are the oligo pools themselves, at around $300 per pool. Another significant expense is the final amplicon sequencing of all replicates (Module 6B), estimated to be less than $2,000. As sequencing is performed using a 150 bp paired-end protocol with standard Illumina adaptors, it does not require custom primers and can be multiplexed with other libraries where appropriate. Another cost worth briefly mentioning is the initial investment in low-passage HAP1 cells from Horizon and HEK 293T from ATCC.

## Quantification and statistical analysis

Much of the data that validates this protocol is available in the Supplementary Information of a biorxiv preprint.[6] We have made examples available in the docs subfolder on the cliPE Github repository ([https://github.com/calhoujd/calhoujd.github.io](https://github.com/calhoujd/calhoujd.github.io)) to help users understand the expected output of Shiny apps and the correct data structure of any necessary input files. The .fastq files can be used to test the Jellyfish commands in Module 7. […]_kmerCount.fa fasta files can be used as example input into the cliPEr_app1_fasta2csv Shiny app. The example_input_CB1_R1.csv and TSC2_e37_kmer_dictionary.csv files can be used as example inputs into the cliPEr_app2_kmers2variants Shiny app. No reverse complementation is required to merge these dataframes and annotate the kmers with variant names. The Book2_e17_3xreps.csv and Book2_e17_4xreps.csv can be used as example inputs into the cliPEr_app3_random_effects_modeling Shiny app. The cliPE pegRNA Designer tool has a built-in default example based on the gene *TSC2* that can be run and users can inspect the different output files.

# Limitations

*CliPE is retrospective in nature*

CliPE generates variants in a region of interest based on variants present in databases of variation, particularly gnomAD and ClinVar. As such, cliPE provides retrospective data based on what variants are present in these databases at the time of designing prime editing reagents. Saturation data such as that generated by other MAVE platforms like deep mutational scanning or saturation genome editing are both retrospective and prospective, at the cost of additional expense, required expertise, and increased hands-on time.

*Variability in prime editing efficiency*

It is important to note that while archetypal epegRNA screening (Module 2) will filter out poorly performing epegRNA architectures, there is still some variability observed within an epegRNA library. There are numerous factors at play, most notably (1) the distance between the PAM site and the edit and (2) the relative proportion of a particular epegRNA within the epegRNA library. The optional QC steps are useful to assess library composition and editing efficiency of each variant. Our prior experience with *TSC2* suggests dropout of ~33% of targeted variants either due to low cloning efficiency or editing efficiency.[6] Design of independent epegRNAs which target overlapping genomic regions is one possibility for alleviating variant dropout, though this would result in an increase in cost per variant overall.

*cliPE requires endogenous expression of the gene of interest in a haploid context*

The method herein is based around prime editing of endogenous loci and therefore requires the following: (1) a cell line which can be co-transfected reasonably efficiently with multiple plasmids, (2) a haploid genome or haploidized locus, and (3) endogenous expression of the gene of interest. Genes endogenously expressed in HAP1 cells are the primary candidates for cliPE and the most likely use case. Genes expressed endogenously in other cell lines may be candidates for cliPE after haploidization of the locus of interest.[2] This can be accomplished using genome engineering tools capable of making large chromosomal deletions.[24] Modifications such as using viral vectors in place of plasmids may be useful for targeting cell lines that are difficult to transfect.

# Troubleshooting

## Problem 1: Incomplete vector digestion (related to multiple cloning steps)

Incomplete DNA digestion is a common issue that causes either cloning failure or reduced cloning efficiency.

## Potential solution:

Any time linearized vector is prepared, it is recommended to run a small (~3 uL) aliquot of vector on a 0.8-1% agarose gel to check for linearization and lack of supercoiled species. It is also recommended to perform an additional ligation reaction with vector only to check for the presence of undigested supercoiled DNA which may confound the subsequent transformation. Minimal outgrowth of vector will be observed in the vector-only ligation transformation if the initial digestion reaction was complete. Significant outgrowth usually suggests incomplete restriction digest of the destination vector. If this is observed, repeat the restriction digest of the backbone vector.

## Problem 2: Failure cloning of nicking gRNAs due to improper design of spacer oligos (related to step 6a)

Cloning of nicking gRNAs follows a classic method widely utilized in genome engineering with CRISPR/Cas9. When this method is done properly, it results in very efficient cloning and hundreds to thousands of clones with the correct plasmid. However, a few common issues can cause issues with this cloning. One of the most common issues is incorrect design of the spacer oligonucleotides.

## Potential solution:

Cloning of nicking gRNAs into BPK1520 follows a similar principle as standard px458/px459 cloning. It is important to order oligonucleotides that conform to the following sticky ends:

ngRNAP_TS: 5' – CACCG – (20 bp spacer) – 3'
ngRNAP_BS: 3' – C – (20 bp spacer) - CAAA – 5.

Remember to use the reverse complement of the spacer sequence for the bottom strand oligo. If the gRNA natively begins with a guanine nucleotide, it is not necessary to add the extra guanine preceding the spacer:

ngRNAP2_TS: 5' – CACC- (20 bp spacer) – 3'
ngRNAP2_BS: 3' – (20 bp spacer)- CAAA – 5.

## Problem 3: Failure cloning of nicking gRNAs due to not diluting preannealed oligos (related to step 6d)

The step in which preannealed spacer oligos are diluted down before the subsequent ligation reaction is often skipped, which often causes failure of the ligation and downstream bacterial transformation.

## Potential solution:

Repeat the cloning protocol with dilution of preannealed spacers.

## Problem 4: Amplification of DNA in the no template control when preparing targeted amplicon sequencing libraries (related to step 10c, 11c, or 11e)

The amplicon sequencing protocol is robust and typically requires minimal optimization or troubleshooting. Rarely, amplicons are observed no template control conditions, suggesting contamination of one of the PCR mastermix components with a trace amount of amplicon.

## Potential solution:

In most cases, discarding DNA grade water aliquots and preparing fresh working stocks (10 uM) forward and reverse primers from 100 uM master stocks corrects this issue.

## Problem 5: HAP1 cells spontaneously become diploid in culture (related to step 5)

Spontaneous diploidization of HAP1 cells has been reported previously.[25] We have observed this phenomenon, especially during the generation of clonal cell lines derivative of parental HAP1s.

## Potential solution:

There are several strategies that can be employed to alleviate this issue. First, it is advised to work with low-passage cells to minimize the number of cells with diploid genomes. Second, it may be necessary to utilize cell sorting to enrich for cells with haploid genomes.[25] Finally, another strategy involves a small molecule, 10-Deacetyl-baccatin-III (DAB), that emerged from an unbiased screen for compounds which selected for haploid cells within mixed cultures.[26] Anecdotally, some labs have reported success with DAB treatment, while others have found little to no effect of drug treatment. Pilot studies with DAB treatment may be useful to determine if treatment enriches for haploid cells.

# Resource availability

*Lead contact*

Further information and requests for resources and reagents should be directed to and will be fulfilled by the lead contact, Jeffrey Calhoun (jeffrey.calhoun@northwestern.edu).

*Technical contact*

Technical questions on executing this protocol should be directed to and will be answered by the technical contact, Jeffrey Calhoun (jeffrey.calhoun@northwestern.edu).

*Materials availability*

Backbone plasmids are available on Addgene (https://www.addgene.org/).

*Data and code availability*

Full datasets will be made available upon publication of the TSC2 MAVE research manuscript. Sufficient excerpts of this data are provided to act as positive controls for cloning steps, prime editing, and data analysis. All Shiny apps, code, examples, etc are available via the cliPE landing page (http://home.clipe-mave.org) and the cliPE github repository (https://github.com/calhoujd/calhoujd.github.io).

## Acknowledgments

This work was sponsored by an American Epilepsy Society Junior Investigator Award (JDC). This work was supported by the Northwestern University – Flow Cytometry Core Facility supported by Cancer Center Support Grant (NCI CA060553). Flow Cytometry Cell Sorting was performed on a BD FACSMelody, purchased through the support of NIH 1S10OD011996-01 and 1S10OD026814-01. We thank David Liu and the Liu lab for sharing their prime editing reagents. The authors would like to acknowledge Addgene for its invaluable service which facilitated this study. We further thank the Plasmidsaurus sequencing team for making quality control and pilot experiments feasible and accessible. Graphical abstract was created in BioRender: Calhoun, J. (2025) https://BioRender.com/c57w266.

## Author contributions

J.D.C., C.G.B., and G.L.C. wrote the manuscript. J.D.C and C.G.B. developed the wet lab methods. J.D.C., C.G.B. and N.B. developed the dry lab methods. N.B. developed the cliPE pegRNA Designer Shiny app. All authors read and approved the manuscript.

## Declaration of interests

The authors declare no competing interests.

## Figures and figure legends

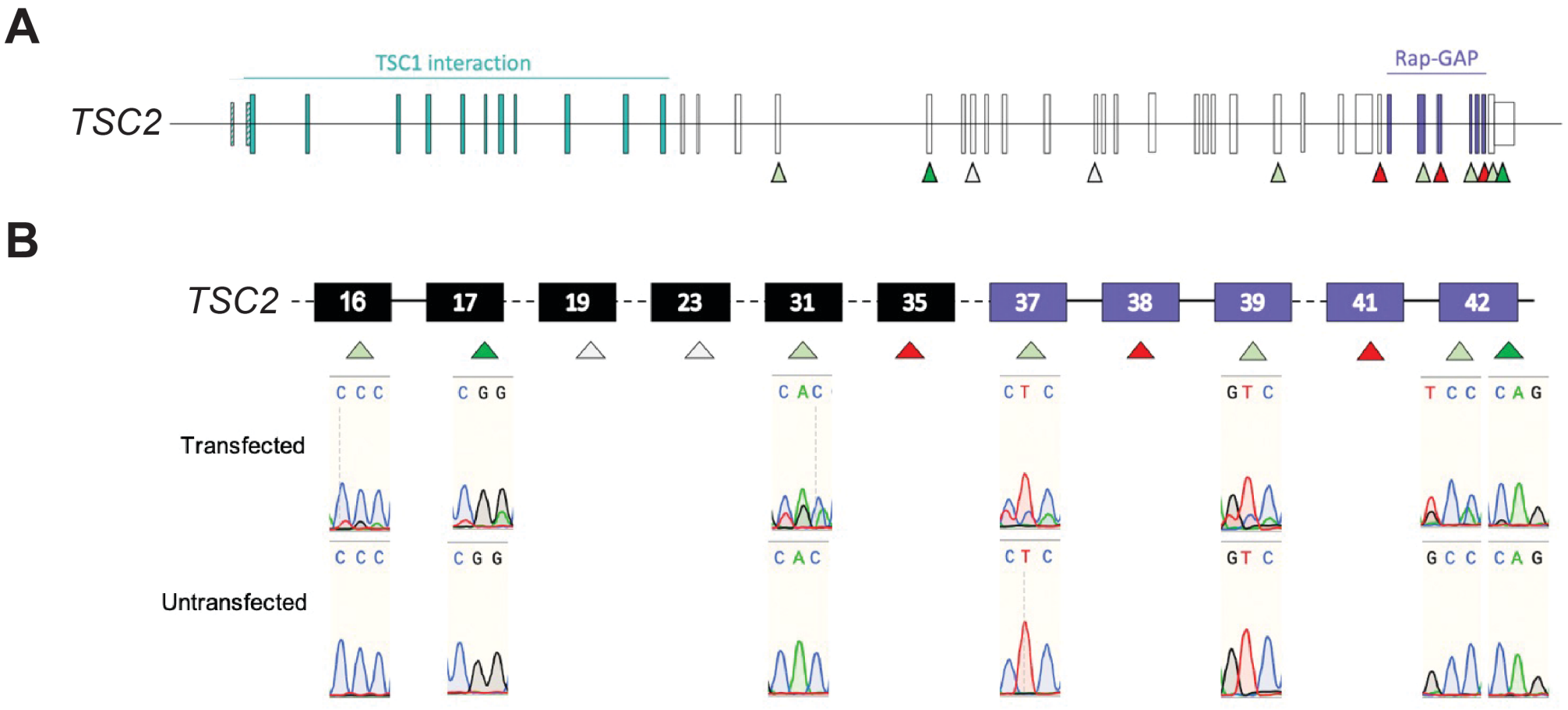

**Figure 1: Example of archetypal epegRNA screen**. (A) Schematic representation of the entire TSC2 gene. Arrowheads indicate targeted regions for screen (B) Excerpt of screen highlighting representative Sanger chromatograms in co-transfected and control untransfected HEK 293T cells. Arrowhead color indicates screening result as follows: light or dark green = efficient editing, red = low/no editing, white = not data not shown. Exons in Rap-GAP domain depicted in purple.

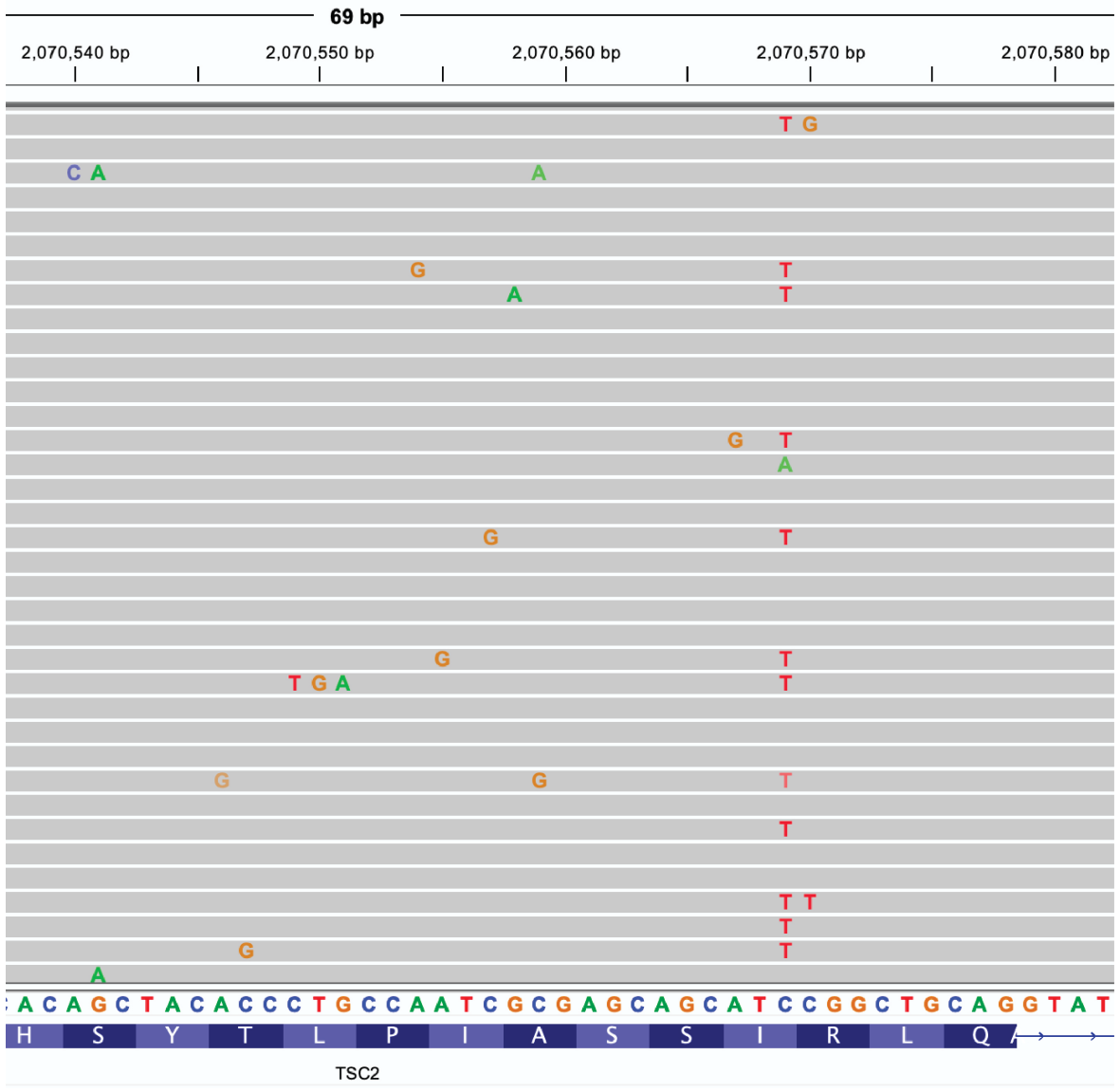

**Figure 2: Example of cliPE editing in HAP1 cells**. Amplicon sequencing was used to validate editing of HAP1 cells in the targeted region of *TSC2* exon 17. Fastq files were aligned to the human hg38 reference and bam files were viewed in IGV software.

# Tables

**Table 1: Outlining truth sets for cliPE experiments.** For most cliPE experiments, the following truth sets are critical to validate the MAVE. For some genes which lack a portion of the truth set, such as *PTEN* or *CHEK2* which have a limited number of BLB variants listed in ClinVar, it is important to consider how the data analysis portion of the workflow may be impacted.

| | **Assay validation truth set** | **Clinical truth set** |
|---|---|---|
| **Negative control** | Synonymous and missense variants present in general population database such as gnomAD | Missense variants classified as benign or likely benign in ClinVar |
| **Positive control** | Premature truncation variants in the targeted regions of the gene | Missense variants classified as pathogenic or likely pathogenic in ClinVar |

**Table 2: Plasmid DNA amounts for co-transfection in epegRNA architecture screen.** We typically use the TurboFectin 8.0 transfection reagent (OriGene #TF81005), but an alternative such as Lipofectamine3000 is suitable if it efficiently transfects HEK 293T cells.

| **Plasmid** | **DNA amount per well of 24-well plate** |
|---|---|
| epegRNA-containing pU6-tevopreq1-GG-acceptor | 75 ng |
| pCMV-PEmax-P2A-GFP | 263 ng |
| pEF1a-hMLH1dn | 132 ng |

**Table 3: Plasmid DNA amounts for co-transfection in epegRNA library selection experiment.** We use the Neon electroporation system which routinely yields 15-30% GFP+ cells. If substituting transfection methods, we recommend optimizing to achieve a minimum of 15% GFP+ cells. Others have reported successful transfection of HAP1s with TurboFectin 8.0 (OriGene #TF81005), although we typically observe significantly reduced efficiency compared to electroporation. Anecdotally, Xfect (Takara; Cat #631317) works well for transfecting small plasmids into HAP1s, but we have not tested this transfection reagent with larger plasmids like the pCMV-PEmax-P2A-GFP.

| **Plasmid** | **DNA amount per 1 million HAP1 cells** |
|---|---|
| epegRNA library | 451 ng |
| Nicking gRNA (in BPK1520) | 180 ng |
| pCMV-PEmax-P2A-GFP | 1579 ng |
| pEF1a-hMLH1dn | 789 ng |